\documentclass{aa}

\usepackage[varg]{txfonts}
\usepackage{graphicx}
\usepackage{adjustbox}

\newcommand{\kms}{\,km\,s$^{-1}$}
\newcommand{\nodata}{.\,.\,.}
\newcommand{\wave}[1]{${\rm \lambda}$#1{\AA}}
\newcommand{\msol}{\,M$_{\sun}$}
\newcommand{\edit}[1]{#1}

\begin{document}

\title{The origin of the planetary nebula M\,1-16}
\subtitle{A morpho-kinematic and chemical analysis}

\author{M.~A. G{\'o}mez-Mu{\~n}oz\inst{\ref{iac},\ref{ull},\ref{iaunam}}
    \and R. V{\'a}zquez\inst{\ref{iaunam}}
    \and L. Sabin\inst{\ref{iaunam}}
    \and L. Olgu{\'\i}n\inst{\ref{unison}}
    \and P.~F. Guill{\'e}n\inst{\ref{iaunam}}
    \and S.~Zavala\inst{\ref{ite}}
    \and R. Michel\inst{\ref{iaunam}}
}

\institute{
Instituto de Astrof{\'\i}sica de Canarias, E-38205 La Laguna, Tenerife, Spain\label{iac}
\and
Departamento de Astrof{\'\i}sica, Universidad de La Laguna, E-38206 La Laguna, Tenerife, Spain\label{ull} \\
\email{magm@iac.es}
\and
Instituto de Astronom\'\i a, Universidad Nacional Aut{\'o}noma de M{\'e}xico, Km 103 Carretera Tijuana-Ensenada, 22860 Ensenada, B. C., Mexico\label{iaunam}
\and
Departamento de Investigaci{\'o}n en F{\'\i}sica, Universidad de Sonora, Blvd. Rosales Esq. L.D. Colosio, Edif. 3H, 83190 Hermosillo, Son. Mexico\label{unison}
\and
Tecnol{\'o}gico Nacional de M{\'e}xico / I.~T. Ensenada, Depto. de Ingenier{\'\i}a en Sistemas Computacionales, C. P. 22780 Ensenada, B. C., Mexico\label{ite}
}

\date{ 2023 / 2023}

\abstract{
We investigated the origin of the Planetary Nebula (PN) M\,1-16 using narrow-band optical imaging,
and high- and low-resolution optical spectra to perform a detailed morpho-kinematic and
chemical studies. M\,1-16 is revealed to be a multipolar PN that predominantly emits in
[\ion{O}{iii}] in the inner part of the nebula and [\ion{N}{ii}] in the lobes.
A novel spectral unsharp masking technique was applied to the position-velocity (PV) maps to
\edit{reveal a set of multiple structures at the centre of M 1-16 spanning radial velocities from
$-$40{\kms} to 20{\kms}, with respect to the systemic velocity.  The morpho-kinematic model indicates that the deprojected velocity of the lobe outflows are $\geq$100{\kms}, and particularly the larger lobes and knots have a deprojected velocity of $\simeq$350{\kms}; the inner ellipsoidal component has a deprojected velocity of $\simeq$29{\kms}. A kinematical age of $\sim$8700\,yr has been obtained from the model assuming a homologous velocity expansion law and a distance of 6.2$\pm$1.9\,kpc}.
The chemical analysis indicates that M\,1-16 is a Type I PN with a 
central star of PN (CSPN) mass in the range of $\simeq0.618-0.713${\msol} and an initial mass for the progenitor star between 
2.0 and 3.0{\msol} (depending on metallicity). An $T_\mathrm{eff}\simeq140\,000$K and $\log(L/{\rm L}_{\sun})$=2.3 was
estimated \edit{using the 
3MdB photoionisation models to reproduce the ionisation. stage of the PN}.
All of these results have led us to suggest 
that M\,1-16 is an evolved PN, contrary to the scenario
of proto-PN suggested in previous studies.
We propose that the mechanism responsible for the morphology of M\,1-16 is related to the 
binary (or multiple star) evolution scenario.
}


\maketitle


\section{Introduction}
\edit{Planetary nebulae (PNe) are astronomical objects representative of one of the latest stages of stellar evolution} of low- and intermediate-mass stars
(0.8--8.0{\msol}). This phase occurs between the asymptotic giant branch (AGB) and white dwarf (WD) stellar phases.
PNe are most known for their complex morphological structures, which are believed to arise during a brief ($\sim$10$^2$--10$^4$~yr) strong mass loss experienced in the post-AGB phase \citep[see the imaging catalogues by][]{lopez2012,sabin2014,parker2016}.
Although the simplest morphologies are well explained with the interactive stellar wind model
and its generalisation \citep[][respectively]{kwok1978,balick1987}, more recent studies
\citep[e.g.][]{soker1992,garciasegura2014,garciasegura2016} demonstrate
that more complex morphologies, such as bipolar PNe, can naturally arise from a binary interaction
via a common-envelope evolution (CEE) \citep[see][for a review]{jones2017}. However,
the complexity of the different morphologies observed in PNe also indicates internal dynamical activity
forming different structural components in the nebula such as jets, collimated outflows, rings, or
knots indicative of different physical processes that are not very well constrained.
The accurate determination of physical and chemical parameters of such PNe, along with
detailed studies of \edit{their morpho-kinematic nature, are useful to} understand their 
formation history \edit{\citep[e.g. Kn~26, IPHASX~J193718.6+202102, NGC~2392, NGC~40, and NGC~6543 by][respectively]{guerrero2013,sabin2021,guerrero2021,rodriguezgonzalez2022,clairmont2022}.}

M\,1-16 ($\alpha_{\rm J2000}=07^{\rm h}37^{\rm m}18{\fs}9$, 
$\delta_{\rm J2000}=-09${\degr}38{\arcmin}47{\farcs}9),
which was first discovered by \citet{minkowski1946}, has a diameter of $\simeq$3\,{\arcsec}
according to a study of optical images by \citet{acker1982}. This measurement was \edit{confirmed by
\citet{kwok1985} using radio interferometry}. An expansion velocity of the ionised gas of $\sim$10{\kms}
was measured in the [\ion{O}{iii}]~$\lambda$5007 emission line by \citet{sabbadin1986}.
\citet{schwarz1992} found that M\,1-16 is a transition
object, often called proto-PN, by means of optical, infrared (IR), and millimetric wavelength observations.
They also found that M1-16 exhibits multiple high-velocity outflows
with shock velocities of around $\sim$300{\kms} \citep{corradi1993} and that the object is surrounded
by a massive envelope of molecular gas. \citet{aspin1993} revealed that M\,1-16  also shows an
extended emission in the near-IR (NIR) bands, mainly produced by vibrational excited
molecular hydrogen (H$_2$), which appears as symmetrically opposed bipolar lobes (as compared with the optical image)
with an extension of $\sim$25{\arcsec}. They argued that these extended emissions of H$_2$
are related to collimated 'jets' with a maximum velocity of 28{\kms} with respect to the central star
(in the plane of the sky). \citet{aspin1993} also found that the effective temperature, $T_{\rm eff}$,
of the central star of PN (CSPN) is $\sim$35\,000~K, as determined by the ratio of the \ion{He}{i} 2.058~$\mu$m
and Br-$\gamma$, which is consistent with a young PN that has recently emerged from the proto-PN phase and is
starting to ionise its surrounding gas.
\edit{Also, high-resolution mapping in the CO $J$=2-1 line of M1-16
revealed a molecular envelope} with a size of 50{\arcsec}, surrounding the ionised central 3{\arcsec} nebula \citep{huggins2000}.  \citet{huggins2000} argued that collimated outflows disrupted the central ionised nebula making cavities, with imprinted CO velocities up to 30{\kms}, along the main axis.

M\,1-16 shows different structural components in the inner regions, close to the central star, and in the outer regions as previously studied by \citet{acker1982}, \citet{schwarz1992}, \citet{aspin1993}, and \citet{corradi1993}.
\citet{Lorenzo2021} also found extended and strong wings in the CO J=9-8 band, indicative of fast collimated outflows. \citet{Guerrero2020} studied the jet nature of M1-16 and found that the distance of the jets, with respect to the CSPN, is $>$0.5\,pc, which is not expected in PNe with high-velocity jets.

In this paper, we present a deep analysis of the morpho-kinematic structure and chemical composition of the PN M\,1-16 which reveal a different evolutionary path from those proposed in previous works. The observational data are described in Sect.~\ref{sec:observational_data}, which includes deep imaging and high-resolution spectroscopy, and the results are presented in Sect.~\ref{sec:results_analysis}. The proposed origin of M\,1-16 is presented in Sect.~\ref{sec:origin_m116} and is supported by our morpho-kinematic model as well as the deep chemical analysis. Finally, the conclusions are presented in Sect.~\ref{sec:conclusions}.

\section{Observational data}
\label{sec:observational_data}

\subsection{Narrow-band optical imaging}

Narrow-band CCD images of M\,1-16 were obtained on 2015 April 4 using the
2.1\,m telescope at the Observatorio Astron{\'o}mico Nacional at the Sierra de San Pedro M{\'a}rtir (OAN-SPM), Mexico.
Observations were carried out using the image mode of the Manchester Echelle Spectrograph
\citep[MES;][]{meaburn2003}. The detector was a 2048$\times$2048 pixel E2V CCD (13.5\,$\mu$m)
with a plate scale of 0{\farcs}351~pix$^{-1}$ (using a 2$\times$2 pixel binning mode). The filters
used to acquire the images were H$\alpha$+[\ion{N}{ii}] ($\lambda_{\rm c}$=6563~{\AA};
$\Delta\lambda$=90~{\AA}) and [\ion{O}{iii}] ($\lambda_{\rm c}$=5007~{\AA}; $\Delta\lambda$=60~{\AA}).
Exposure times were 1800\,s for both filters with a seeing of $\simeq$2\,{\arcsec}.

In addition, narrow-band CCD direct images of M\,1-16 were obtained on 2021 April 13 using
the 0.84~m telescope ($f$/15) at the OAN-SPM. The CCD camera MEXMAN was used to obtain deep images of M\,1-16
in the H$\alpha$ ($\lambda_{\rm c}$=6565~\AA, $\Delta\lambda$=11~\AA) and [\ion{N}{ii}] 
($\lambda_{\rm c}$=6583~\AA, $\Delta\lambda$=10~\AA) narrow bands. The detector was a 2024$\times$2024 E2V CCD
with a size of 15\,$\mu$m\,pix$^{-1}$ (resulting in a nominal plate scale of 0.22\,{\arcsec}\,pix$^{-1}$).
Exposure times were 1200\,s and 600\,s in H$\alpha$ and [\ion{N}{ii}], respectively, using a binning of 2$\times$2 pixels.
Seeing was around 1.9\,{\arcsec} during the observations. The images were processed using standard {\sc iraf}\footnote{The Image Reduction and Analysis Facility ({\sc iraf}) is a collection of software written at the National Optical Astronomy Observatory (NOAO) geared towards the reduction of astronomical images in pixel array form.} routines. The MEXMAN images provide a different view of M\,1-16 in the H$\alpha$ and [\ion{N}{ii}] emission lines separately, contrary to the images obtained with the MES instrument in which the observations are from the combined H$\alpha$+[\ion{N}{ii}] emission lines.

\subsection{High-resolution optical spectroscopy}

Long-slit high-resolution optical spectra were also acquired with the MES instrument. We set a slit width of 150\,$\mu$m (1.9\,{\arcsec})
with a fixed length of {6.5\,{\arcmin}}. Spectra were obtained during a five-day observational 
campaign that began on 2015 April 4. Observations were carried out using the 2$\times$2
binning mode (plate scale of 0.351\,{\arcsec}\,pix$^{-1}$). The spectra were centred around the
H$\alpha$ filter with $\Delta\lambda=90$\,{\AA} to isolate the 87$^{\rm th}$ order
(0.1\,{\AA}~pix$^{-1}$ spectral scale). The exposure time for all of the spectra was set to 1800\,s.
The seeing was $\simeq1${\arcsec} as measured from a full width at half maximum (FWHM) fit taken from the field stars' continuum along the spatial axis of the spectra.

All the spectra were processed with standard techniques for long-slit spectroscopy using
{\sc iraf}. The wavelength calibration was performed using a ThAr arc lamp to
an accuracy of $\pm$1{\kms}. The FWHM of the arc-lamp emission lines was measured to be $\simeq$12{\kms}.

\subsection{Boller \& Chivens long-slit low-resolution optical spectroscopy}

Low-resolution optical spectra for the M\,1-16 were acquired using the Boller \& Chivens
spectrograph mounted in the 2.1\,m telescope at OAN-SPM, during one observing run on
2015 November 19. An E2V CCD with a 2048$\times$2048
pixel array and plate scale of 1.8\,{\arcsec}\,pix$^{-1}$ (1$\times$2 binning mode) was used as
a detector. A spectral resolution of $\simeq$5.0{\AA} (FWHM), as judged by the arc calibration lamp
spectrum, was obtained by using the 400\,lines\,mm$^{-1}$ grating in conjunction with the 2\,{\arcsec} wide
slit, covering a spectral range of 4100--7600\,{\AA}.

Different exposure times were used in order to enhance different structural components in  
M\,1-16.
We used the median of three 180\,s exposure spectra in order to study the bright central part of
the nebula and ten spectra (three of 300\,s, three of 180\,s, two of 1800\,s, one of 900\,s, and one of 60\,s) were co-added, leading to a total of 6000\,s, to study the
faintest regions along the main axis. 

The spectra were reduced using standard procedures for long-slit spectroscopy under
{\sc iraf} software. The calibration was performed using the standard star Feige 34.
The stellar component was extracted and separated from the surrounding nebular emission when possible.

\section{Results}
\label{sec:results_analysis}

\subsection{Morphology}

\begin{figure*}
    \centering
    \includegraphics[width=\textwidth]{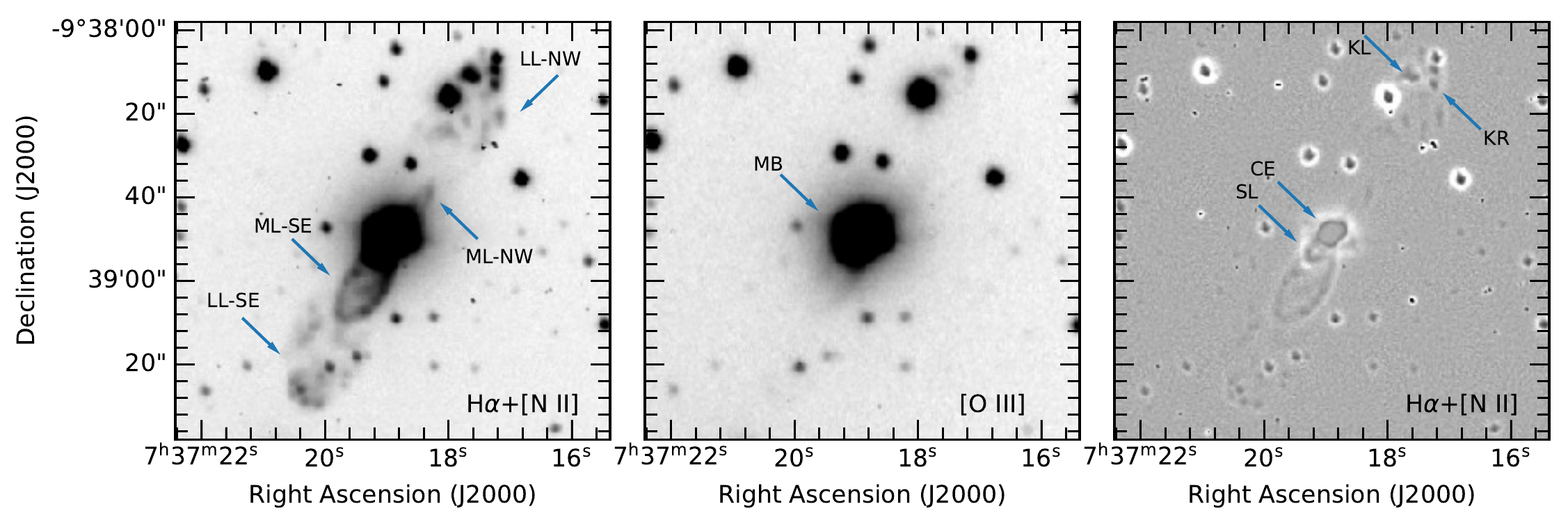}
    \includegraphics[width=0.70\textwidth]{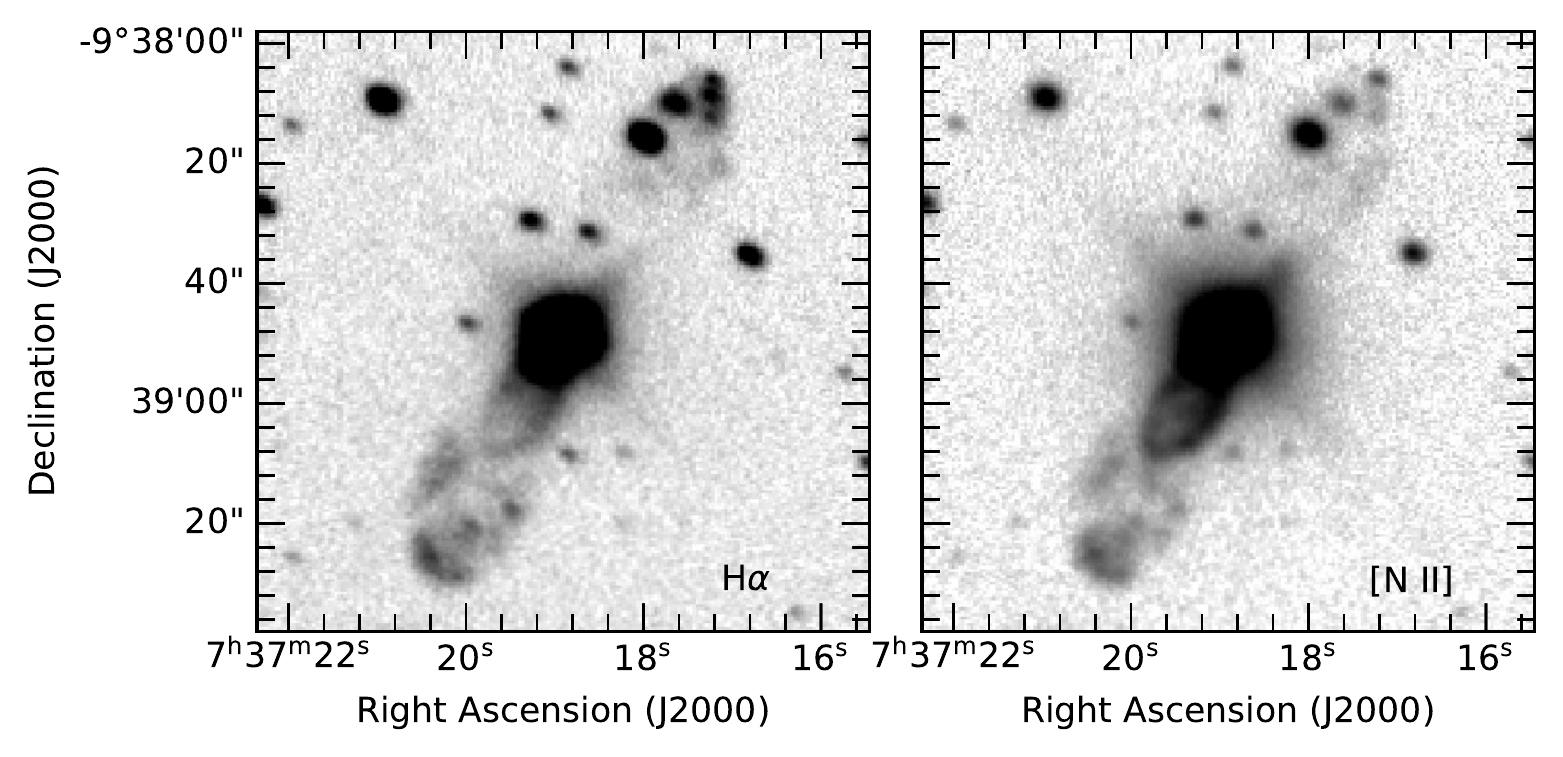}
    \caption{Narrow band optical images of M\,1-16 a seen in different emission lines. Top panel: M\,1-16 as seen in H$\alpha$+[\ion{N}{ii}] (left), [\ion{O}{iii}] (middle), and unsharp masked H$\alpha$+[\ion{N}{ii}] (right)
    images obtained with MES in image mode. Bottom panel: Separated H$\alpha$ (left) and [\ion{N}{ii}] (right) images obtained
    with the MEXMAN camera. Arbitrary contrast scales were set to all images to enhance the different features seen in the nebula.
    North is up and east is to the left in all of the images.}
    \label{fig:m116_optical_images}
\end{figure*}

Our deep images, H$\alpha$+[\ion{N}{ii}]{\wave{6584}}, H$\alpha$, [\ion{N}{ii}]\wave{6584}, and [\ion{O}{iii}]{\wave{5007}},
obtained with MES in image mode and with the MEXMAN camera, show
the complex seemingly multi-polar morphology of M\,1-16.
Initially, we identified two pairs of lobes or outflows coming out from
the central bright region of M\,1-16 (Fig.~\ref{fig:m116_optical_images}).
One pair of large lobes (LL), previously discovered by \citet{schwarz1992}
and \citet{aspin1993}, is seen with a position angle (PA) on the plane of
the sky of $\simeq-$31{\degr}, and they are labelled in the Fig.~\ref{fig:m116_optical_images}
as LL-SE and LL-NW (the SE and NW lobe, respectively). We measured their angular extension to be $\simeq$47\,{\arcsec} and $\simeq$51\,{\arcsec}, with respect to the bright central region or main body (MB), for the LL-SE and LL-NW, respectively. 
A similar value of 90\,{\arcsec} for the full extent of the LL was measured by \citet{schwarz1992}. 
A pair of Middle-sized Lobes (ML) are seen close to the MB, whose size in the sky is
$\simeq$24\,{\arcsec} and $\simeq$15\,{\arcsec} for the ML located to the SE (ML-SE) and NW (ML-NW),
respectively, with a PA$\simeq-$41{\degr}. The ML-SE correspond to the loop
seen by \citet{aspin1993} in the NIR which primarily traces the emission of the vibrationally excited H$_{\rm 2}$.\\ %

Finally, we used the unsharp-masking technique to enhance the structural components of the MB in M\,1-16, allowing us to identify new features
(see the right image at the top panel of Fig.~\ref{fig:m116_optical_images}).
Our images show a smaller lobe (SL-SE) inside the MB, which has a size of $\sim$8{\arcsec} and a PA$\simeq -$39{\degr}. We did not observe any NW counterpart in our images.
 In this figure we also found a central ellipse (CE) $\simeq9\times7$\,{\arcsec} with its major axis at PA=$-$55{\degr}.

\begin{figure}
    \centering
    \includegraphics[width=1.\columnwidth]{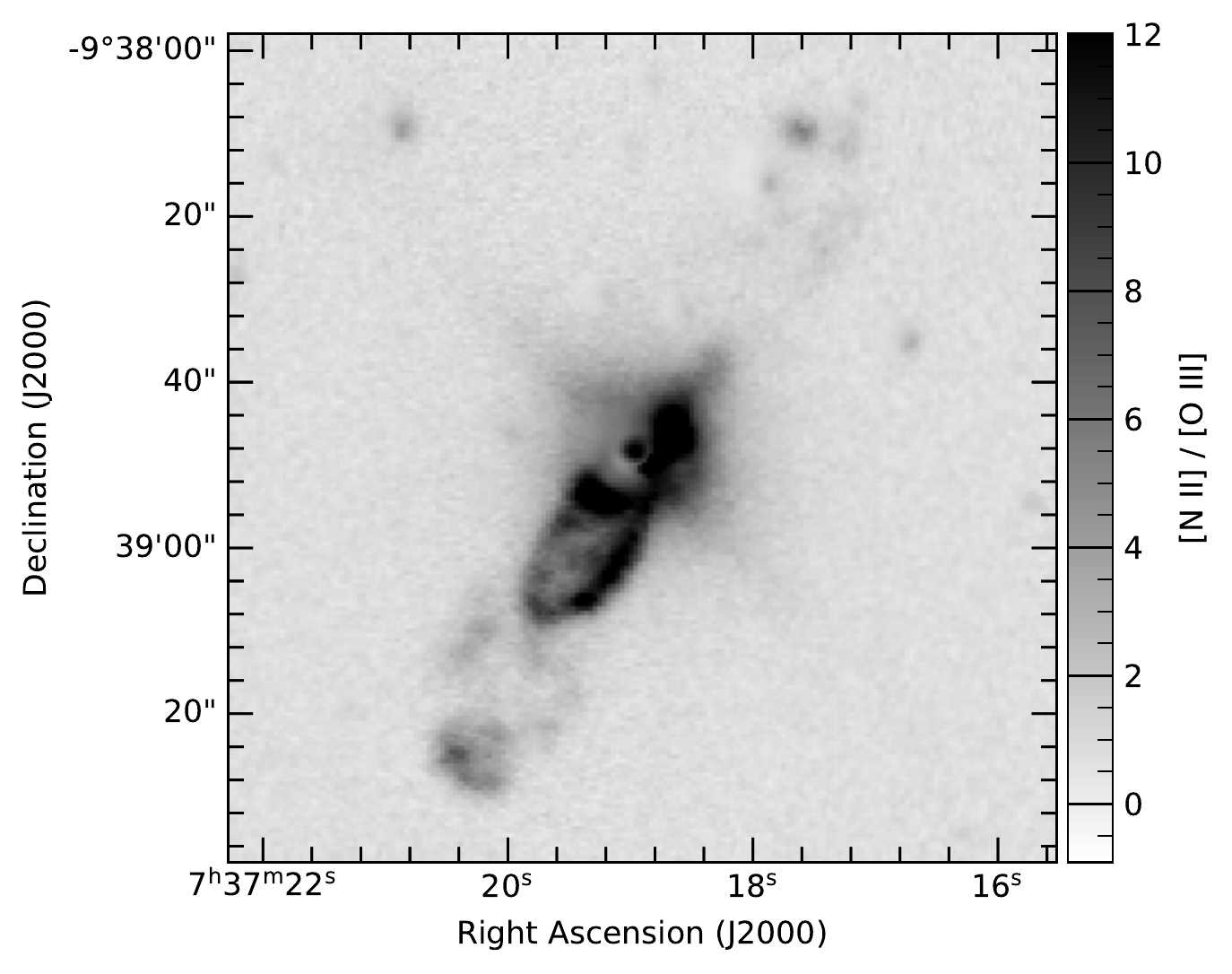}
    \caption{Grey-scale image of the [\ion{N}{ii}]/[\ion{O}{iii}] of M\,1-16 where dark values indicate a dominant [\ion{N}{ii}] emission line over [\ion{O}{iii}].
    \label{fig:ratio_HaNii_Oiii}}
\end{figure}

Figure~\ref{fig:ratio_HaNii_Oiii} shows the image ratio [\ion{N}{ii}]/[\ion{O}{iii}] of M\,1-16.  The [\ion{O}{iii}] nebular emission is slightly stronger in the central part of M\,1-16 (in white in the figure), faint in the ML (mostly ML-SE), and barely visible in the large lobes (see the middle image in the top panel of Fig.~\ref{fig:m116_optical_images}).
The [\ion{N}{ii}] nebular emission, which is a known tracer
of low-ionisation structures \citep{corradi1996}, is concentrated in the lobes showing different
structural components such as over-densities and knots (darker components in Fig.~\ref{fig:ratio_HaNii_Oiii}).

\subsection{Kinematics}
\label{subsec:kinematics}

\begin{figure}
\includegraphics[width=\columnwidth]{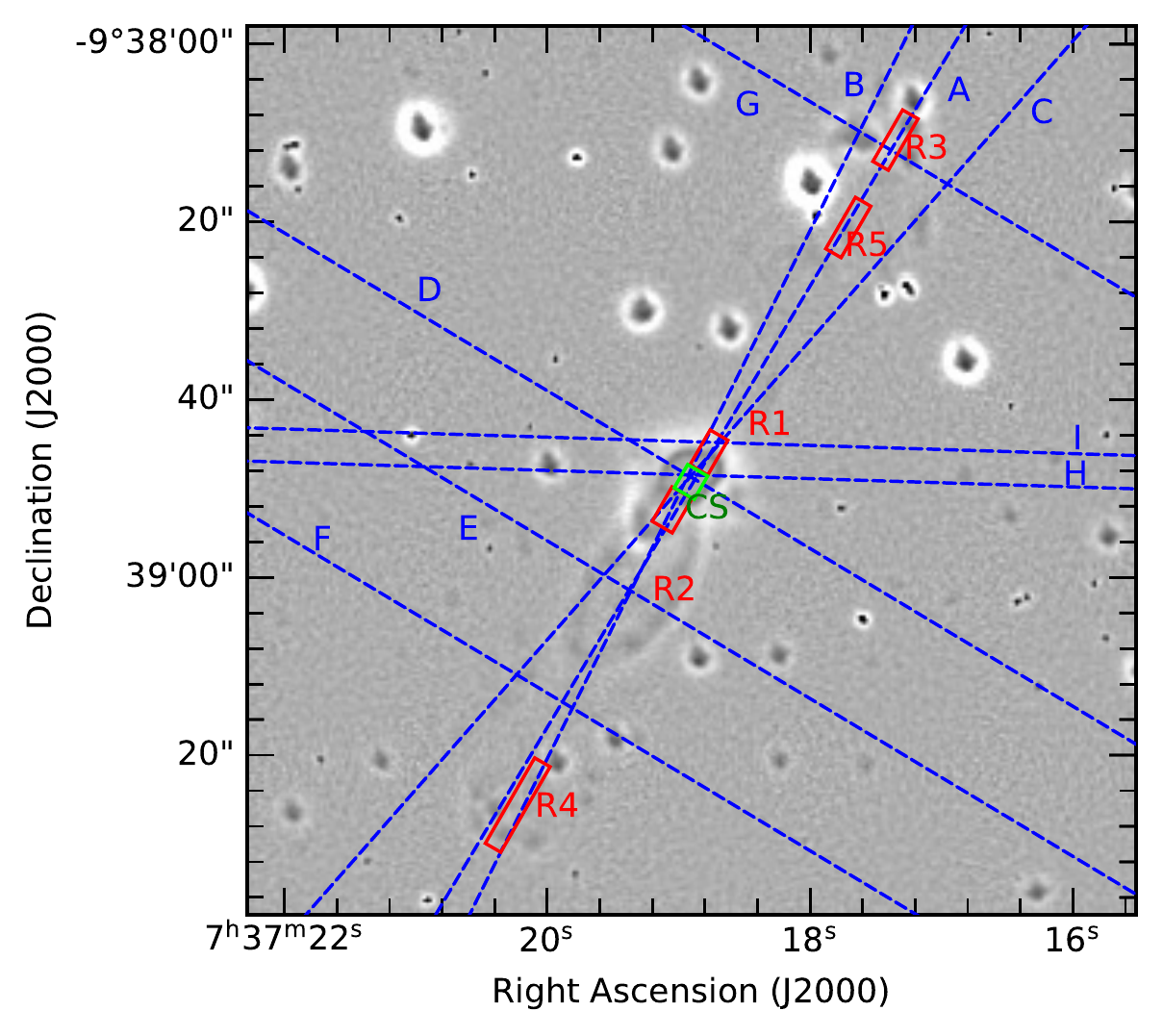}
\caption{M\,1-16 as seen in H$\alpha$+[\ion{N}{ii}] unsharp-masked images taken with MES in image mode. The MES 
and the B\&Ch slits are overlaid in blue dashed lines and red boxes, respectively. The CS region is shown as a green box. \label{fig:mes_boler_slits}}
\end{figure}

Position-velocity (PV) diagrams corresponding to the slit paths in Fig.~\ref{fig:mes_boler_slits},
were obtained and they are shown in Figs.~\ref{fig:mez_nii_spectra} and \ref{fig:mez_ha_spectra} (see Appendix~\ref{apendice_PVs}),
labelled from A to I. The position angle (PA) for slits A, B, and C correspond to $-30\degr$, $-25\degr$, and $-40\degr$, respectively. The other slits have  PA$=60\degr$ (D to G) and PA$=90\degr$ (H and I). Spectra corresponding to slits passing
through the central star (A, B, C, D, and H) were used to measure the systemic velocity of the shell, 
which was estimated as 
$V_{\rm sys}^{\rm LSR}=48\pm1$\,km\,s$^{-1}$. 
A visual inspection of PV maps A, B, and C (Figs.~\ref{fig:mez_nii_spectra} and \ref{fig:mez_ha_spectra}) shows that they are 
dominated by the central emission, which exhibits a widening at the position of the central 
star (also shown in PV maps D and H). The [\ion{N}{ii}] PV maps A, B, and C, also show an 
evident protrusion leaving the central region towards the SE, along the slit, and that is blue-shifted. 
This protrusion continues up to 20\,{\arcsec} (or even more in PV map B), increasing its 
radial velocity to $\simeq120$\,km\,s$^{-1}$. This feature is not visible in H$\alpha$ (just a marginal 
emission in B). PV map C also shows a fainter antisymmetrical protrusion towards the NW.
The PV maps and measurements of the observed radial velocities are with
respect to the aforementioned systemic velocity, $V_\mathrm{sys}^\mathrm{LSR}$.

\begin{figure*}
    \centering
    \includegraphics[width=1.\textwidth]{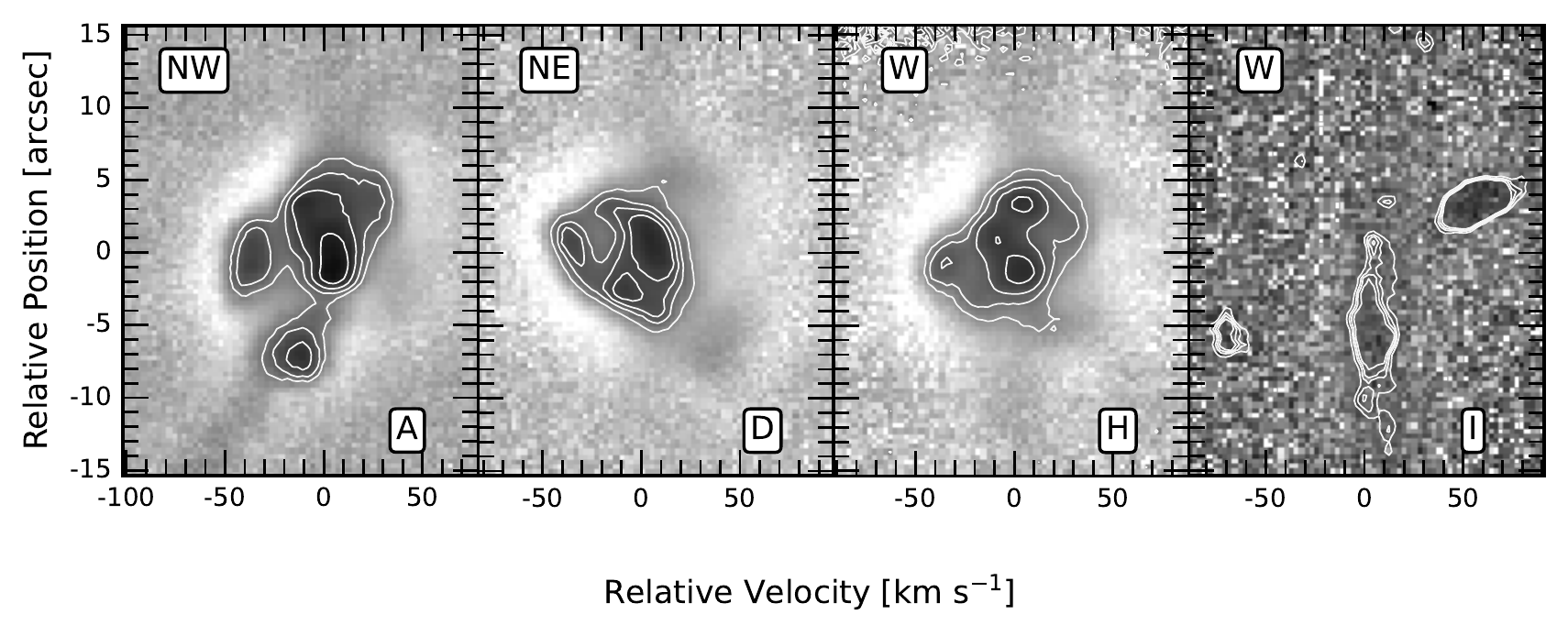}
    \caption{Grey-scale image and contour plot of the [\ion{N}{ii}] emission line PV maps B, D, H, and I of the central region of M\,1-16 using the spectral unsharp masking technique (see Appendix~\ref{apendice_sum}). Both the image and the contour plot are in logarithmic representation with an arbitrary colourbar and levels, respectively.
    \label{fig:umaskN2}}
\end{figure*}

On the other hand, PV map E shows a deformed ellipse with an expansion velocity of 
$v_{\rm exp}\simeq30$\,km\,s$^{-1}$ centred at $\simeq-80$\,km\,s$^{-1}$ with respect to the 
systemic velocity. Additionally, PV maps F and G show double flat structures, spanning 
radial velocities of $140$ and $250\pm1$\,km\,s$^{-1}$, respectively, with the first one being centred at $-80$ and the last one at 120$\pm1$\,km\,s$^{-1}$.
The PV map I marginally shows emission at the systemic velocity spanning 20\,{\arcsec}, as well as two faint knots at +50 and $-50$\,km\,s$^{-1}$. 
 PV maps at H$\alpha$ A, B, C, D, and H are dominated by the central emission, also showing
 the \ion{He}{ii}$\lambda6560$\,{\AA} emission line at the centre, whereas E, F, G, and I are a 
 faint version of the corresponding PV maps in [\ion{N}{ii}]. 
Finally, a spectral logarithmic unsharp masking technique (see 
Appendix~\ref{apendice_sum}) was applied on the [\ion{N}{ii}] PV maps A, D, H, and I showing 
that, at the centre, the spectral line is divided into some components in velocity, 
spanning \edit{radial velocities} from $-40$ to $+20\pm1$\,km\,s$^{-1}$ (see contour-grey PV maps in 
Fig.~\ref{fig:umaskN2}).

\subsection{Chemical and physical conditions}

We identified and extracted five nebular zones along the major axis of the nebula. In addition, we extracted the area corresponding to the central star. The regions are identified from R1 to R5 and CS, and are shown in Fig.~\ref{fig:mes_boler_slits}
as red boxes. We obtained the median of the spectra extracted from the inner regions, R1 and R2, to study the bright central part of M\,1-16 and we added the multiple spectra taken for R3, R4, and R5 to study its faintest regions.

The calculations of the physical conditions, abundances, and respective uncertainties  were performed with the code {\sc pyneb} \citep{Luridiana2015}, which is a tool dedicated to nebular analysis. The atomic data used to determine the ionic abundances are those described in \citet{Morisset2020}. 

Line fluxes for each extracted region were measured with the `splot' task by fitting a Gaussian function to each line. The error budget included read-out noise and photon noise produced by both emission lines and the sky at the line position. However, uncertainties arising from wavelength calibration, flux calibration, or emission line fitting were not considered. To calculate the uncertainties on the line fluxes, we followed the independent and random errors approach \citep{Taylor1997}. The error propagation (for all the parameters) was based on a Monte Carlo approach similarly to the work by \citet{Sabin2022} for the PN PC~22.

From the observed line intensities and their associated errors, {\sc pyneb} calculates the logarithmic extinction $c({\rm H\beta})$, the electronic density  $N_{\rm e}$, and temperature $T_{\rm e}$ in an iterative way, starting as a first step with a theoretical ratio of H$\alpha$/H$\beta$=2.85, $T_{\rm e}=10^4$~K, and $N_{\rm e}=10^3$~cm$^{-3}$, assuming a case B recombination \citep[][]{osterbrock2006}.
After convergence, the final values of $c({\rm H\beta})$, $N_{\rm e}$, and $T_{\rm e}$ were obtained.
The $T_{\rm e}$ was calculated using the [\ion{O}{iii}](\wave{5007}$+$\wave{4959})/\wave{4963}
and [\ion{N}{ii}](\wave{6548}$+$\wave{6583})/\wave{5755} ratios, for high- and low-excitation regions, respectively.
In the case of $N_{\rm e}$, we used the [\ion{S}{ii}]\wave{6716}/\wave{6731},
[\ion{Ar}{iv}]\wave{4711}/\wave{4740}, and [\ion{Cl}{iii}]\wave{5537}/\wave{5518} ratios,
when available.

The first (top) half of Table~\ref{tab:line_fluxes} shows the intensity of the emission lines obtained for the different regions studied here.
The second (bottom) half shows the extinction value, $c({\rm H\beta})$, and the physical conditions, electronic density ($N_{\rm e}$), and temperature ($T_{\rm e}$),
obtained for the different regions.
The emission lines were dereddened with the estimated value of $c({\rm H\beta})$ and using the extinction law of \citet{ccm89}.
The values of $N_{\rm e}$ and $T_{\rm e}$  could only be estimated simultaneously for CS and regions R1, R2, R3, and R4, only. The interpretation of these results is shown in Sec.~\ref{subsec:chemistry}.  \\

\begin{table*}
    \centering
    \caption{Dereddened emission-line fluxes (normalised to H$\beta$ = 100) and derived physical parameters.} 
    \label{tab:line_fluxes}
    \begin{tabular}{lrrrrrrr}
    \hline
\hline
Ion                               &  Line & CS & R1 & R2 & R3  & R4 & R5 \\
\hline
H$\delta$+\ion{He}{ii}            & $\lambda$4101  &    28.4$\pm$  0.4 &      88.9$\pm$     4.2 &    38.8$\pm$     1.3 &  \nodata &  \nodata &  \nodata \\ 
H$\gamma$                         & $\lambda$4340  &    50.6$\pm$  0.3 &      84.5$\pm$     4.5 &    54.1$\pm$     1.5 &    37.2$\pm$     1.7 &    65.6$\pm$     7.6 &  \nodata \\ 
$[$\ion{O}{iii}$]$                & $\lambda$4363  &    17.6$\pm$  0.2 &      26.4$\pm$     2.7 &    17.4$\pm$     1.5 &  \nodata &  \nodata &  \nodata \\ 
\ion{He}{i}                       & $\lambda$4388  &     1.2$\pm$  0.1 &    \nodata &  \nodata &  \nodata &  \nodata &  \nodata \\ 
\ion{He}{i}                       & $\lambda$4471  &     5.6$\pm$  0.1 &    \nodata &     5.1$\pm$     1.0 &  \nodata &  \nodata &  \nodata \\ 
\ion{N}{iii}+\ion{O}{ii}          & $\lambda$4640  &     4.1$\pm$  0.1 &    \nodata &  \nodata &  \nodata &  \nodata &  \nodata \\ 
\ion{He}{ii}                      & $\lambda$4686  &    25.7$\pm$  0.1 &    \nodata &     8.9$\pm$     0.7 &  \nodata &  \nodata &  \nodata \\ 
\ion{He}{i}+$[$\ion{Ar}{iv}$]$    & $\lambda$4711  &     3.7$\pm$  0.1 &    \nodata &  \nodata &  \nodata &  \nodata &  \nodata \\ 
$[$\ion{Ne}{iv}$]$                & $\lambda$4724  &     0.4$\pm$  0.1 &    \nodata &  \nodata &  \nodata &  \nodata &  \nodata \\ 
$[$\ion{Ar}{iv}$]$                & $\lambda$4740  &     3.3$\pm$  0.1 &    \nodata &  \nodata &  \nodata &  \nodata &  \nodata \\ 
H$\beta$                          & $\lambda$4861  &   100.0$\pm$  0.3 &     100.0$\pm$     3.0 &   100.0$\pm$     0.9 &   100.0$\pm$     1.2 &   100.0$\pm$     7.6 &   100.0$\pm$     5.6 \\ 
$[$\ion{O}{iii}$]$                & $\lambda$4959  &   399.4$\pm$  0.9 &     328.0$\pm$     7.4 &   304.5$\pm$     2.2 &    22.0$\pm$     0.7 &    13.9$\pm$     2.6 &    20.4$\pm$     4.2 \\ 
$[$\ion{O}{iii}$]$                & $\lambda$5007  &  1190.8$\pm$  2.5 &     916.9$\pm$    20.0 &   890.9$\pm$     6.1 &    60.3$\pm$     0.9 &    45.3$\pm$     2.8 &    77.1$\pm$     4.7 \\ 
$[$\ion{N}{i}$]$                  & $\lambda$5199  &     6.9$\pm$  0.1 &      34.1$\pm$     2.0 &    28.0$\pm$     0.5 &    43.3$\pm$     0.7 &    63.3$\pm$     3.0 &    96.2$\pm$     4.9 \\ 
\ion{He}{ii}                      & $\lambda$5411  &     2.1$\pm$  0.1 &    \nodata &  \nodata &  \nodata &  \nodata &  \nodata \\ 
$[$\ion{Cl}{iii}$]$               & $\lambda$5517  &     0.6$\pm$  0.1 &    \nodata &  \nodata &  \nodata &  \nodata &  \nodata \\ 
$[$\ion{Cl}{iii}$]$               & $\lambda$5537  &     0.7$\pm$  0.1 &    \nodata &  \nodata &  \nodata &  \nodata &  \nodata \\ 
$[$\ion{N}{ii}$]$                 & $\lambda$5755  &    16.9$\pm$  0.1 &      17.6$\pm$     1.6 &    24.1$\pm$     0.5 &  \nodata &  14.6 $\pm$   2.0 & 13.0 $\pm$ 3.0 \\ 
\ion{He}{i}                       & $\lambda$5876  &    16.7$\pm$  0.1 &      12.9$\pm$     1.7 &    16.3$\pm$     0.5 &     6.1$\pm$     1.0 &  \nodata &  \nodata \\ 
$[$\ion{O}{i}$]$                  & $\lambda$6300  &    14.5$\pm$  0.1 &      30.5$\pm$     2.0 &    37.3$\pm$     0.5 &    41.7$\pm$     0.8 &    43.9$\pm$     2.9 &    47.5$\pm$    3.9 \\ 
$[$\ion{S}{iii}$]$+\ion{He}{ii}   & $\lambda$6312  &     0.9$\pm$  0.1 &    \nodata &  \nodata &  \nodata &  \nodata &  \nodata \\ 
$[$\ion{O}{i}$]$                  & $\lambda$6363  &     4.9$\pm$  0.1 &      9.5 $\pm$     1.3 &    12.5$\pm$     0.5 &    13.8$\pm$     0.6 &    14.7$\pm$ 2.4 &    14.0 $\pm$     2.3 \\ 
\ion{He}{ii}                      & $\lambda$6407  &     0.2$\pm$  0.1 &    \nodata &  \nodata &  \nodata &  \nodata &  \nodata \\ 
$[$\ion{Ar}{v}$]$                 & $\lambda$6435  &     0.4$\pm$  0.1 &    \nodata &  \nodata &  \nodata &  \nodata &  \nodata \\ 
$[$\ion{N}{ii}$]$                 & $\lambda$6548  &   233.2$\pm$  0.7 &     268.2$\pm$     8.6 &   378.8$\pm$     3.7 &   190.2$\pm$     2.4 &   153.4$\pm$    6.3 &   126.1 $\pm$    7.5 \\ 
H$\alpha$                         & $\lambda$6563  &   283.0$\pm$  0.9 &     282.8$\pm$     9.0 &   283.6$\pm$     2.8 &   286.2$\pm$     3.5 &   280.3$\pm$    11.4 &   279.7$\pm$    6.0 \\ 
$[$\ion{N}{ii}$]$                 & $\lambda$6583  &   707.3$\pm$  2.1 &     730.6$\pm$    22.9 &  1132.3$\pm$    11.1 &   580.6$\pm$    7.1 &   467.8 $\pm$    18.8 &   392.2$\pm$    22.4 \\ 
\ion{He}{i}                       & $\lambda$6678  &     4.2$\pm$  0.1 &       5.9$\pm$     2.0 &     5.9$\pm$     0.4 &  \nodata &  \nodata &  \nodata \\ 
$[$\ion{S}{ii}$]$                 & $\lambda$6716  &     5.2$\pm$  0.1 &     23.7 $\pm$     1.2 &    34.2$\pm$     0.6 &    43.5$\pm$     0.8 &    59.8$\pm$    3.0 &    72.9 $\pm$    4.8 \\ 
$[$\ion{S}{ii}$]$                 & $\lambda$6731  &     7.0$\pm$  0.1 &     19.4 $\pm$     1.4 &    31.1$\pm$     0.5 &    43.2$\pm$     0.8 &    47.6$\pm$     2.8 &    50.8$\pm$    3.8 \\ 
$[$\ion{Ar}{v}$]$                 & $\lambda$7006  &     0.8$\pm$  0.1 &    \nodata &  \nodata &  \nodata &  \nodata &  \nodata \\ 
\ion{He}{i}                       & $\lambda$7065  &     6.9$\pm$  0.1 &    \nodata &     7.6$\pm$     0.5 &  \nodata &  \nodata &  \nodata \\ 
$[$\ion{Ar}{iii}$]$               & $\lambda$7136  &    21.1$\pm$  0.1 &    \nodata &    21.2$\pm$     0.6 &  \nodata &  \nodata &  \nodata \\ 
\ion{He}{i}                       & $\lambda$7281  &     0.9$\pm$  0.1 &    \nodata &  \nodata &  \nodata &  \nodata &  \nodata \\ 
$[$\ion{O}{ii}$]$                 & $\lambda$7320  &    11.9$\pm$  0.1 &    \nodata &    10.7$\pm$     0.8 &     6.7$\pm$     0.9 &    16.1$\pm$     3.3 &  \nodata \\ 
$[$\ion{O}{ii}$]$                 & $\lambda$7330  &     7.9$\pm$  0.1 &    \nodata &     7.9$\pm$     0.5 &    13.3 $\pm$    0.9 &     4.1$\pm$     2.6 &  \nodata \\
\hline
c(H$\beta$)                       &       &    0.88$\pm$ 0.01 &    \nodata &   0.59 $\pm$   0.01  &   0.60 $\pm$    0.01 & 0.69   $\pm$    0.04 &    0.80 $\pm$    0.06 \\
T$_e$($[$\ion{N}{ii}$]$)          &       &   12\,100$\pm$ 100  &     12\,300$\pm$     850 &   11\,600$\pm$    100  &  \nodata &   14200$\pm$ 1300 & 15000 $\pm$ 2200 \\
T$_e$($[$\ion{O}{iii}$]$)         &       &   13\,300$\pm$ 100  &     18\,300$\pm$    1270 &   15\,000$\pm$    600  &     100$\pm$ \nodata &  \nodata &  \nodata \\
N$_e$($[$\ion{S}{ii}$]$)          &       &    2300$\pm$ 130  &      276$\pm$     120 &     500$\pm$    50  &   610  $\pm$    240  &      230 $\pm$ 93  &     $<$1   \\
N$_e$($[$\ion{Cl}{iii}$]$)        &       &    4100$\pm$ 1000 &    \nodata &  \nodata &  \nodata &  \nodata &  \nodata \\ 
\hline
    \end{tabular}
\end{table*}

\begin{table*}
    \centering
    \caption{Derived ionic abundances.}
    \label{tab:ionic_ab}
    \begin{tabular}{lrrrrrr}
    \hline
\hline
Ion.      &                       CS               &   R1                & R2                  & R3                  & R4                  &    R5         \\
\hline
He$^{+}$ ($\times$ $10^{1}$)  &   1.225 $\pm$ 0.036 &  1.225 $\pm$ 0.273  &  1.510 $\pm$ 0.068  & 0.437 $\pm$ 0.052   & \nodata             & \nodata              \\ 
He$^{2+}$ ($\times$ $10^{2}$) &   3.029 $\pm$ 0.243 &  \nodata            &  0.771 $\pm$ 0.061  & \nodata             & \nodata             & \nodata               \\ 
O$^{0}$  ($\times$ $10^{5}$)  &   1.520 $\pm$ 0.013 &  2.825 $\pm$ 0.494  &  4.300 $\pm$ 0.138  & 8.136 $\pm$ 0.152   &   2.675 $\pm$ 0.781 &  2.839 $\pm$   1.590 \\ 
O$^{+}$ ($\times$ $10^{4}$)   &   1.100 $\pm$ 0.016 &  \nodata            &  2.032 $\pm$ 0.140  & 33.990 $\pm$ 3.243  & 0.838 $\pm$ 0.266   & \nodata               \\ 
O$^{2+}$  ($\times$ $10^{4}$) &   1.739 $\pm$ 0.019 &  0.635 $\pm$ 0.085  &  0.968 $\pm$ 0.100  & 21.640 $\pm$ 0.031  &   0.053 $\pm$ 0.017 &   0.086 $\pm$   0.042 \\ 
N$^{0}$  ($\times$ $10^{6}$)  &   5.905 $\pm$ 0.116 & 21.450 $\pm$ 4.039  & 21.370 $\pm$ 0.723  &  33.040 $\pm$ 1.355 &  24.260 $\pm$ 8.710 &  40.370 $\pm$  25.870 \\ 
N$^{+}$  ($\times$ $10^{5}$)  &   9.262 $\pm$ 0.054 &  9.441 $\pm$ 1.087  & 15.980 $\pm$ 0.300  &  11.320 $\pm$ 0.114 &   4.259 $\pm$ 0.873 &   3.573 $\pm$   1.274 \\ 
S$^{+}$  ($\times$ $10^{7}$)  &   2.706 $\pm$ 0.036 &  6.954 $\pm$ 0.833  & 12.330 $\pm$ 0.262  &  22.050 $\pm$ 0.602 &  12.800 $\pm$ 2.416 &  14.580 $\pm$   4.816 \\ 
S$^{2+}$  ($\times$ $10^{7}$) &   7.113 $\pm$ 0.568 &  \nodata            & \nodata             & \nodata             & \nodata             & \nodata               \\ 
Cl$^{2+}$ ($\times$ $10^{8}$) &   5.253 $\pm$ 0.273 &  \nodata            & \nodata             & \nodata             & \nodata             & \nodata               \\ 
Ar$^{2+}$ ($\times$ $10^{6}$) &   1.161 $\pm$ 0.007 &  \nodata            &  1.253 $\pm$ 0.040  & \nodata             & \nodata             & \nodata               \\ 
Ar$^{3+}$ ($\times$ $10^{7}$) &   4.328 $\pm$ 0.105 &  \nodata            & \nodata             & \nodata             & \nodata             & \nodata               \\ 
Ar$^{4+}$ ($\times$ $10^{8}$) &   8.358 $\pm$ 0.430 &  \nodata            & \nodata             & \nodata             & \nodata             & \nodata               \\ 
Ne$^{5+}$ ($\times$ $10^{5}$) &   6.378 $\pm$ 1.005 &  \nodata            & \nodata             & \nodata             & \nodata             & \nodata               \\
\hline
    \end{tabular}
    \tablefoot{Ionic abundances were calculated using T$_e$($[$\ion{N}{ii}$]$) and N$_e$($[$\ion{S}{ii}$]$) for low ionisation potential (IP) and  T$_e$($[$\ion{O}{iii}$]$) and N$_e$($[$\ion{S}{ii}$]$) for high IP ions when available. Low IP ions were considered when IP$<$30~eV.}
\end{table*}

\begin{table}
    \centering
    \caption{Derived elemental abundances.}
    \label{tab:elemental_ab}
    \begin{adjustbox}{max width=\columnwidth}
    \begin{tabular}{lrrrr}
    \hline
    \hline
X/H             &    CS~~~~~~~~~        &   R1~~~~~~~~ &  R2~~~~~~~~~       &    Solar\tablefootmark{a}          \\
\hline
He/H    ($\times10^{1}$)  &  1.526 $\pm$  0.002 &    1.125 $\pm$ 0.273  & 1.585  $\pm$  0.069   &  0.840\\
O/H     ($\times10^{4}$)  &  3.236 $\pm$  0.043 &               \nodata & 3.077  $\pm$  0.175   &  5.370\\
N/H     ($\times10^{4}$)  &  2.556 $\pm$  0.029 &               \nodata & 2.528  $\pm$  0.103   &  0.724 \\
S/H     ($\times10^{6}$)  &  1.427 $\pm$  0.047 &               \nodata & 3.626  $\pm$  0.142   & 14.500 \\
Cl/H    ($\times10^{8}$)  &  7.906 $\pm$  0.346 &               \nodata &               \nodata &  1.780 \\
Ar/H    ($\times10^{6}$)  &  1.436 $\pm$  0.017 &               \nodata & 1.487  $\pm$  0.062   &  3.160 \\
\hline
log(O/H)                 &$-$3.489 $\pm$ 0.002  &               \nodata & $-$3.512 $\pm$ 0.024  & \\
log(N/H) &                $-$3.592 $\pm$ 0.007  &               \nodata & $-$3.597 $\pm$ 0.066  &  \\
log(N/O) &                 $-$0.103 $\pm$ 0.003 &               \nodata & $-$0.085 $\pm$ 0.026     & $-$0.870 \\
\hline
    \end{tabular}
    \end{adjustbox}
\tablefoot{\tablefoottext{a}{Solar abundances from \citet{lodders2010}.}}
\end{table}

\section{The nature of M1-16}
\label{sec:origin_m116}

\subsection{Morpho-kinematic structure}
\label{subsec:kinematic_structure}

\begin{figure}
    \centering
    \includegraphics[width=0.7\columnwidth]{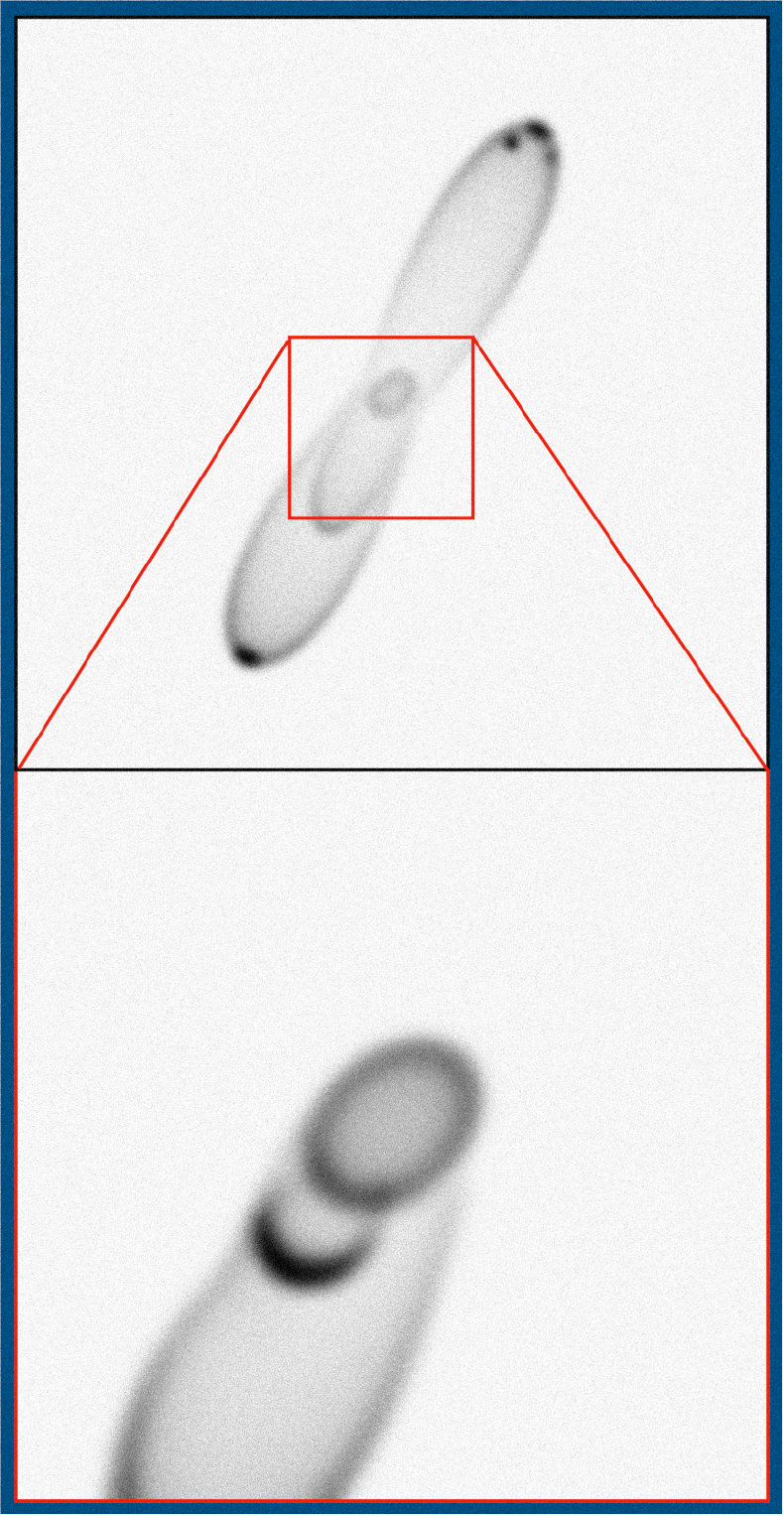}
    \caption{{\sc ShapeX} reconstruction of M\,1-16 based on all PV diagrams. The top panel shows the fill model and the bottom panel shows the inner-central part of the model.}
    \label{fig:shapeX_fullmodel}
\end{figure}

Morphology and kinematics of M\,1-16 were modelled using {\sc ShapeX}, a software that specialises in morpho-kinematic reconstructions of gaseous nebulae \citep{steffen2011}.
Our analysis determined that the structures that provide the best simultaneous fit to both the imaging and spectra data are: (1) two large lobes, (2) one medium semi-lobe, (3) one (apparently compact) small semi-lobe, (4) one central ellipsoidal structure, (5) two distinct bipolar blobs, and (6) two knots in the northern lobe (see Fig.~\ref{fig:shapeX_fullmodel}).

\begin{figure*}
    \centering
    \includegraphics[width=.8\textwidth]{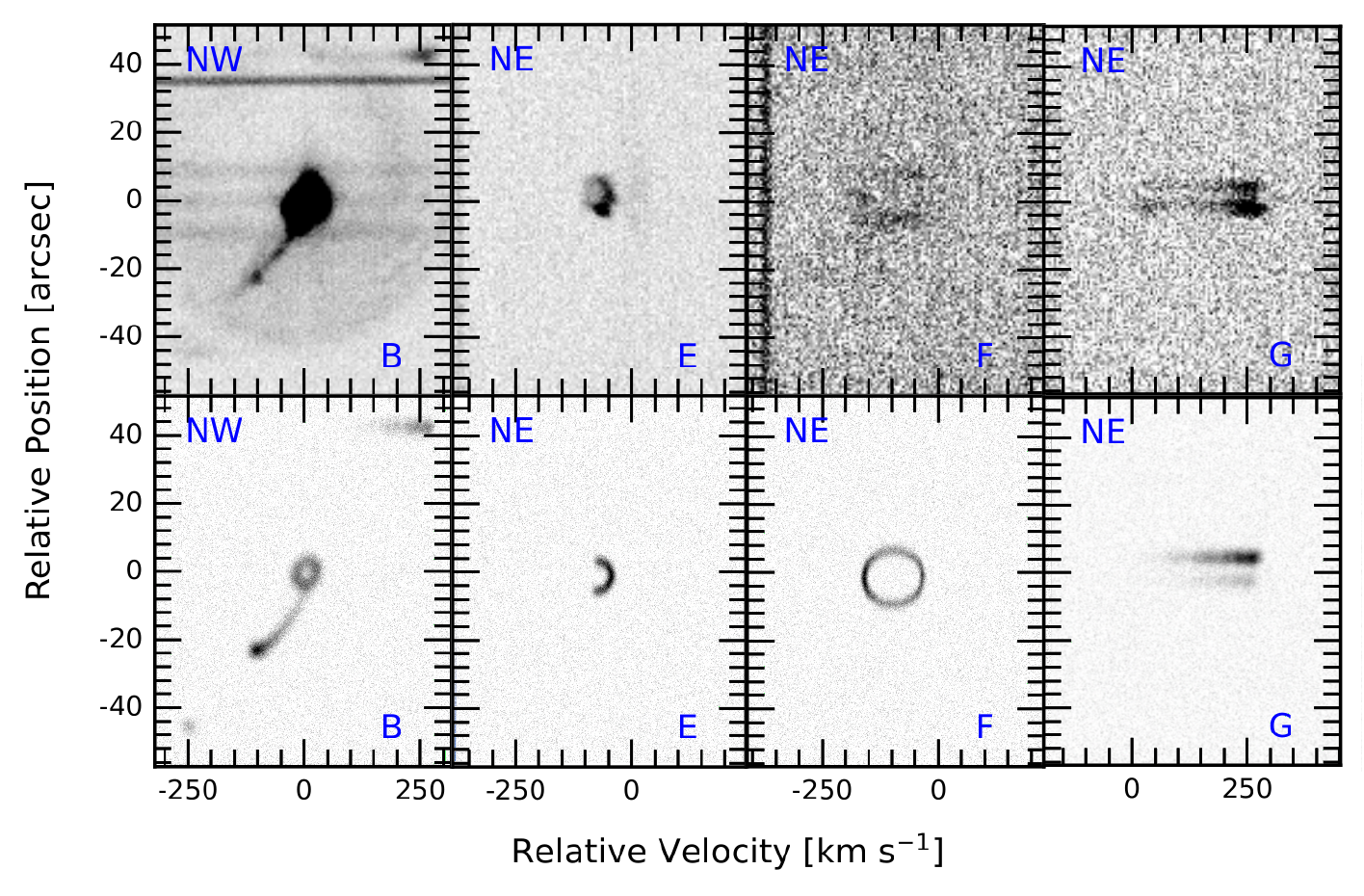}
    \caption{Comparison of the observed PV maps in the [\ion{O}{iii}] emission line for B, E, F, and G slits (top panel) and the {\sc ShapeX} synthetic spectra obtained from the adopted model (bottom panel).
    \label{fig:PV_model}}
\end{figure*}

\begin{figure}
    \centering
    \includegraphics[width=1.\columnwidth]{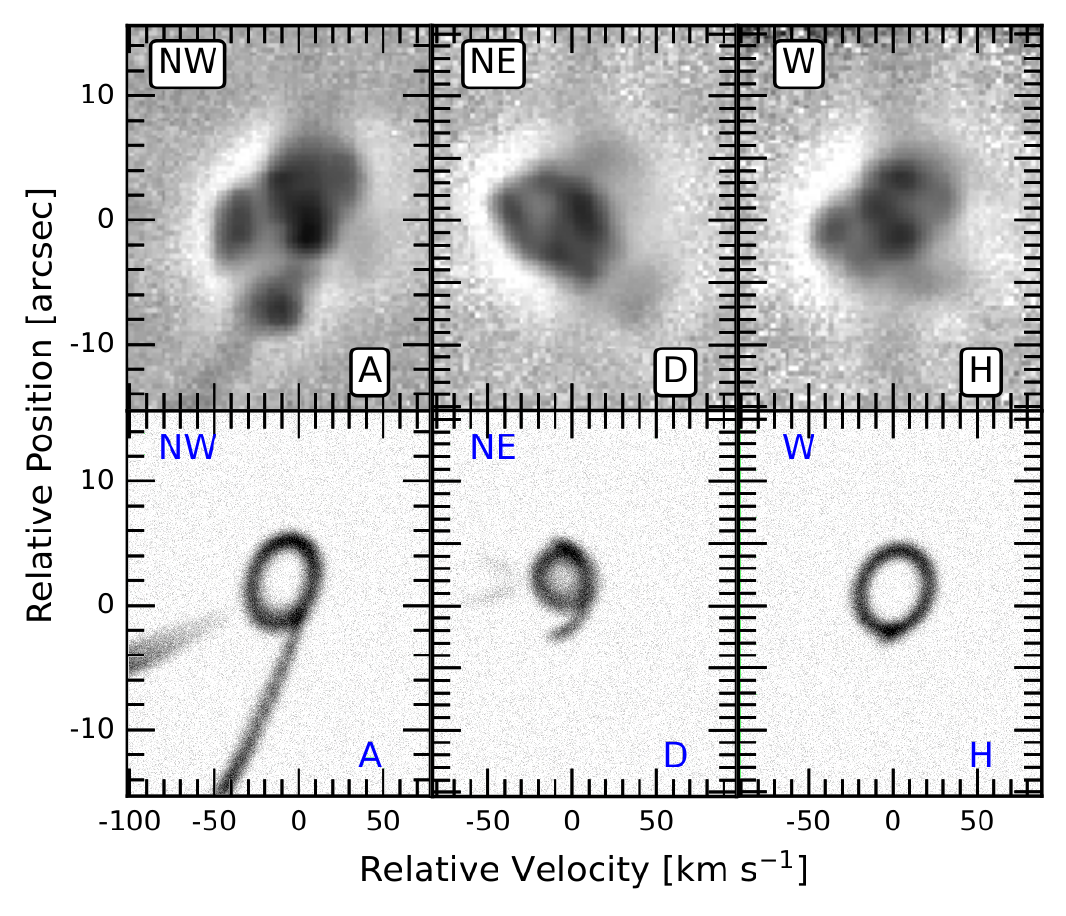}
    \caption{Similar to Fig.~\ref{fig:PV_model} but for A, D, and H slits of the central part of M1-16.
    \label{fig:PV_core_model}}
\end{figure}

To determine the size, shape, and velocity law of each morphological structure, a visual fitting process was performed. The velocity law of a structure establishes the relationship between deprojected velocity and size or distance in a 3D vectorial basis. We used homologous velocity laws in which $\vec V \propto \vec r$. Figures~\ref{fig:PV_model} and \ref{fig:PV_core_model} display the comparison between these structures and the observations. The two blobs correspond exactly with the tips of the bipolar lobes, so we do not mention them afterwards. A summary of the main parameters of each structure, including their kinematic age and relationship with features labelled in Fig.~\ref{fig:m116_optical_images}, can be found in Table~\ref{tab:shapex}.

\begin{table}
    \centering
    \caption{Morphological structures.}
    \label{tab:shapex}
    \resizebox{1.\columnwidth}{!}{
    \begin{tabular}{lrccccc}
    \hline
    \hline
Label & Structure  & $\theta$ & PA & $i$ & \edit{$V_{\rm dep}$} & $\tau_{\rm k}$\\
 &  & ($\pm 2$\,{\arcsec}) & ($\pm 2 \degr$) & ($\pm 2 \degr$) &
 ($\pm 1$\,km\,s$^{-1}$) & ($\pm 1800$ years)\tablefootmark{a}    \\
\hline
LL-NW, LL-SE & Large lobes & 52 & $-$29 &  118 & 354 & \edit{4300} \\
ML-SE & Medium semi-lobe &   35 & $-$27 &  140 & 140 & \edit{7300} \\
SL & Small semi-lobe     &   37 & $-$38 &  170 & 124 & \edit{8700} \\
CE & Central ellipsoid   &    4 & $-$48 &  110 &  29 & \edit{4000} \\
LK & Left knot (north)   &   47 & -25 & 115 & 350 & \edit{3900} \\
RK & Right knot (north)  &   49 & -34 & 116 & 350 & \edit{4000} \\
\hline
    \end{tabular}
    }
    \tablefoot{
    	For the label and geometry of the structures refer to Fig.~\ref{fig:m116_optical_images}. Parameters: $\theta$ is the polar or semi-major axis for lobes,
    	semi-major axis for the central ellipsoid, and angular distance for the knots.
    	PA 	and $i$ denote the direction (position angle and inclination with respect to the line of sight) of the radial vector for the knots and the semi-major
    	axis for the rest of the structures. Furthermore, $V_{\rm dep}$ represents the deprojected polar velocity for structures and deprojected velocity for the
    	knots, with respect to the systemic velocity, while $\tau_{\rm k}$ indicates the kinematical age of each structure based on its radial velocity and
    	physical size, using the distance to M\,1-16 of \edit{6.2$\pm$1.9\,kpc \citep{frew2016}} and the formulation from \citet{Guillen2013}. \\
    	\tablefoottext{a}{Mean value of the uncertainty (from 1200 up to 2700 as minimum and maximum values, respectively).}
            }
\end{table}

It is noteworthy that both knots were modelled using the same velocity law as the large lobes and placed close to the surface. \edit{As a result, the deprojected polar velocity of the large lobes and the deprojected velocity of the knots from the central star exhibit good agreement. Based on this model, it can be inferred that the three structures share the same kinematic age},
 suggesting that knots are part of or were impacted by the expansion of the bipolar outflow that created the large lobes. However, we also modelled the knots as small cylinders \edit{($1\,\text{{\arcsec}}$ in diameter and length)}, with asymmetric density \edit{and an internal expansion velocity law (that is, with respect to the centre of each knot) of $V_{\text{exp}} \simeq 100\,\text{km s}^{-1} \times r$}. Our goal was to model bow-shock-like structures in order to reproduce the morpho-kinematical structures seen on the PV maps. The small size of the cylinders takes into account the spatially non-resolved bow shocks.

By considering the kinematic ages of all structures, it appears that the mass-loss history of this object may have progressed as follows: (1) the first collimated ejection of the SL stream (which may have been bipolar), (2) the second collimated ejection of the ML flow (which may have also been bipolar), (3) the bipolar collimated ejection of the LL, which interact with the LK and RK nodes, and finally (4) the ejection of the CE flow.
The sequence of events leading to the observed structure may have been caused by a binary or multiple system, as similar scenarios have been suggested in the past \citep[see][for a review]{jones2017}.
The unusual morpho-kinematic structures of the SL and ML could result from the partial ionisation of bipolar structures that are not fully exposed to radiation, possibly due to the obstructing effect of a warped disk surrounding the central star. A careful analysis of the unsharp masking processed image (see the top-right panel of Fig.~\ref{fig:m116_optical_images}) reveals the presence of additional, faint lobes and ejections, some of which may be related to the semi-lobes. The rapid expansion of the knots, spreading at \edit{radial velocities} of more than 200\,km\,s$^{-1}$ with the tips reaching close to 250 km\,s$^{-1}$, could be explained by the effect of high-velocity ejected clumps colliding with the shell or the bipolar outflow interacting with previously ejected mass-loaded knots. This could also account for the slightly elevated temperature in [\ion{N}{ii}] relative to other regions.

High-velocity outflows are a common feature in PNe, especially in those with bipolar or point-symmetric shapes. These outflows are thought to be driven by collimated winds from the CSPN or its binary companion, and they interact with the slow-moving circumstellar envelope ejected during the AGB phase. One example of such a PN is M 3-38, which has a bipolar morphology and a high-velocity component defined as a jet \citep{RechyGarcia2022}. This PN has been proposed to be the result of a CEE, where the primary star engulfed a main-sequence companion with a 17\,day orbital period. Then, the CEE phase may produce a bipolar proto-PN with fast outflows and a circumbinary disk \citep{GarciaSegura2021}. A similar scenario has been suggested for other PNe with high-velocity outflows, such as MyCn18, KjPn8, and also M~1-16 \citep[][and references therein]{Guerrero2020}. However, the origin and evolution of these outflows are still poorly understood and require further observations and modelling.

It is crucial to note that, from the comparison between observations and the model depicted in
Fig.~\ref{fig:PV_core_model}, it can be inferred that the nature of the core of M\,1-16 is clumpy.
While it is true that these clumps may generally adhere to the model's pattern, their
non-uniformity suggests a discontinuous ejection of material.

\subsection{Chemical and physical parameters}
\label{subsec:chemistry}

Table~\ref{tab:line_fluxes} shows the variation of the logarithmic extinction across the nebula. The most internal, and slightly more obscured regions' CS, shows $c({\rm H\beta})$=0.88 with a mean value of $\simeq$0.74 (taking into account the central region R2), whereas external regions R3, R4, and R5 have a similar mean value of $\simeq$0.70.

The electronic density calculated with the low-ionisation species, [\ion{S}{ii}], shows fluctuations along the central region CS, R1, R2, and R3 regions, with $N_{\rm e}$ being a factor $\sim$4 lower in R1, R2, and R3 compared to the CS. This implies that the collisional excitation rate is less important in R1, R2, and R3 with respect to the CS. The estimation of $N_{\rm e}$  with the high-ionisation species, that is [\ion{Cl}{iii}] and [\ion{Ar}{iv}], could only be performed in the internal region CS.

According to Fig.~\ref{fig:mes_boler_slits}, we are currently measuring the $N_{\rm e}$ from the SL-SE and the MB in R2 and the CS, respectively.
The value of $N_{\rm e}$([\ion{S}{ii}])=2300~cm$^{-3}$ of the CS is in agreement with the value obtained by \citet{corradi1993} as measured
in the central region of M\,1-16, and much higher than the one estimated by \citet{schwarz1992}.
The $N_{\rm e}$=610~cm$^{-3}$ for the LL-NW, as measured
in R3, however, is higher than the value obtained by \citet{corradi1993} of 200~cm$^{-3}$ (for that low regime; see Sec.~\ref{subsec:kinematics}). The $T_{\rm e}$, estimated from [\ion{N}{ii}] and [\ion{O}{iii}] for low- and high-ionisation regions, respectively, does not show fluctuation and a mean value of $\sim$14~kK is obtained.

Estimated ionic and elemental abundances are given in Table~\ref{tab:ionic_ab} and \ref{tab:elemental_ab}, respectively. To estimate the elemental abundances, we used the ionisation correction factors (ICF) of \citet{DelgadoInglada2014}. 
When compared to solar abundances derived by \citet{lodders2010} (see Table~\ref{tab:elemental_ab}),
M\,1-16 is clearly N and Cl overabundant with (N/H)$\simeq$3.53\,(N/H)$_{\sun}$
and (Cl/H)$\simeq$4.44\,(Cl/H)$_{\sun}$, respectively.
\edit{On the contrary, M~1-16 shows a sub-solar abundance} in O, S, and Ar with
(O/H)$\simeq$0.60\,(O/H)$_\sun$, (S/H)$\simeq$0.098\,(S/H)$_\sun$, and
(Ar/H)$\simeq$0.45\,(Ar/H)$_\sun$, respectively. In the case of He, the variation is only 1.82
with respect to the solar abundance.

The elemental abundance results indicate that M\,1-16 is a Type-I PN
because of the $\log({\rm N/O})=-$0.103 and He/H=0.153 obtained for the CS
\citep[defined as $\log({\rm N/O})>-$0.3 and He/H$\geq$0.125;][]{stanghellini2010}.
We also estimated the CSPN mass by using method three described in \citet{maciel2010},
yielding \edit{$\sim$0.74{\msol}}. According to \citet{millerbertolami2016},
such a CSPN mass indicates that M\,1-16 evolved from a progenitor
star with a mass between 3 and 4{\msol} for $Z$=0.02 or 2.5{\msol} for $Z$=0.001.

\citet{schwarz1992} and \citet{aspin1993} argued that M\,1-16 is a young object, the so-called proto-PN,
and the latter estimated a $T_\mathrm{eff}$=35\,000~K by means of the ratio between
\ion{He}{i}\,2.058~$\mu$m and Br$\gamma$. However, 
the presence of high excitation ions, such as \ion{He}{ii}\wave{4686}, [\ion{O}{iii}]\wave{5007}, and
[\ion{Ar}{v}]\wave{6435}, in the optical spectra, indicate a $T_\mathrm{eff}$ of at
least 50\,000--60\,000~K.
We used the 3MdB \citep{Morisset2015}, a database of photionisation models made 
with the CLOUDY code \citep[v17.01;][]{Ferland2017}, to estimate the $T_\mathrm{eff}$ of the CSPN of M~1-16. We looked
for models made with the theoretical atmosphere models from \citet{Rauch2003}
and the ones that  simultaneously reproduce the log(\ion{He}{ii}\wave{4686}/H$\beta$), 
log([\ion{O}{iii}] \wave{5007}/[\ion{N}{ii}] \wave{6583}), 
log([\ion{Ar}{v}] \wave{7006}/[\ion{Ar}{iv}] \wave{4740}), and
log([\ion{O}{iii}] \wave{5007}/[\ion{O}{i}] \wave{6300}). A total
of 13 models were identified to fit the observed ionisation stage of M~1-16,
corresponding to a $T_\mathrm{eff}\simeq$140\,000~K and a
$\log(L/L_\sun)\simeq2.3$. Therefore, according to the theoretically evolutionary post-AGB
tracks of \citet{millerbertolami2016},
this sets a progenitor mass between 2.0--3.0\,M{\sun} and a mass for the CSPN between 0.618--0.713\,M{\sun}
depending on the metallicity (comparable with the estimated value for the CSPN mass using the
log(N/O)).

\section{Conclusions}
\label{sec:conclusions}

\edit{From the kinematic study, we derived a systemic velocity
for M~1-16 of $V_\mathrm{sys}^\mathrm{LSR}=48\pm1${\kms} and a deprojected expansion velocity of 
$\simeq$30{\kms} for the brightest lobe, ML-SE}, as obtained from the PV maps presented here. A detailed
morpho-kinematic model was build using the computational model {\sc shape} fitting our
data and improving our interpretation.
The model is composed by two pairs of lobes, corresponding to SL and ML, and one medium
semi-lobe, one central ellipsoidal structure, and two pairs of knots located to the N at the tip
of the LL-NW.
According to the model, M\,1-16 \edit{first ejected the SL stream with a
deprojected velocity of 125{\kms}} ($i$=10{\degr}) followed
by a second collimated ejection of the ML flow at 140{\kms} ($i$=40{\degr}). Then, the
bipolar collimated ejection of the LL occurred and interacted with the LK and RK
nodes. Finally, the ejection of CE flow occurred. A binary system or multiple system at the
centre of M\,1-16 is proposed to explain the sequence of mass-loss events in this nebula,
as similar scenarios have been observed in other cases \citep[see][and references therein]{jones2017}. The unusual morpho-kinematic structures of the
SL and ML could result from the partial ionisation of bipolar
structures that are not fully exposed to radiation, possibly due to
the obstructing effect of a warped disk surrounding the central
star.

The chemical analysis along with the kinematical age obtained from the morpho-kinematic model
demonstrated that M\,1-16 \edit{may be an evolved PN instead of a proto-PN as previously suggested}.
The presence of high ionisation lines, such
as [\ion{O}{ii}], [\ion{Cl}{iii}], and [\ion{He}{ii}], and the chemical abundances indicate that
the CSPN have a mass of $\sim$0.74{\msol} and evolved from a progenitor star with a mass between
2.5 and 4.0{\msol} (depending on the metallicity); a $T_\mathrm{eff}\simeq$140\,000~K and a $\log(L/\mathrm{L}_\sun)\simeq$2.3 were estimated
by using the \edit{3MdB photoionisation models} in conjunction with theoretical post-AGB evolutionary models \edit{indicating a CSPN mass in the range of 0.618--0.713{\msol} and a progenitor mass
between 2.0--3.0{\msol}}. 

\section*{Acknowledgements}

MAGM acknowledges support from the ACIISI, Gobierno de Canarias and the European Regional Development Fund (ERDF) under grant with reference PROID2020010051 as well as from the State Research Agency (AEI) of the Spanish Ministry of Science and Innovation (MICINN) under grant PID2020-115758GB-I00.
MAGM also acknowledges the postdoctoral fellowship granted by the Instituto de Astronom{\'{\i}}a of the
Universidad Nacional Aut{\'o}noma de M{\'e}xico.
This work was supported by UNAM-PAPIIT IN106720 grant. LS acknowledges support from grant UNAM-PAPIIT IN110122. SZ acknowledges support from the ITE-UNAM agreement 1500-479-3-V-04, and from O.Velázquez-DISC/ITE.  This work is based upon observations carried out at the Observatorio Astron{\'o}mico Nacional on the Sierra San Pedro M{\'a}rtir (OAN-SPM), Baja California, Mexico.
We thank the daytime and night support staff at the OAN-SPM for facilitating and
helping obtain our observations. In particular to the telescope operators
Mr. Gustavo Melgoza (also known as `Tiky') and Mr. Felipe Montalvo.
This research made use of APLpy, an open-source plotting package for Python \citep{aplpy}.
We thank the referee for they comments that improved the manuscript.


\bibliographystyle{aa} 
\bibliography{main_v2} 


\begin{appendix}
\section{Position-velocity maps}
\label{apendice_PVs}

\begin{figure*}
    \centering
    \includegraphics[width=1.0\textwidth]{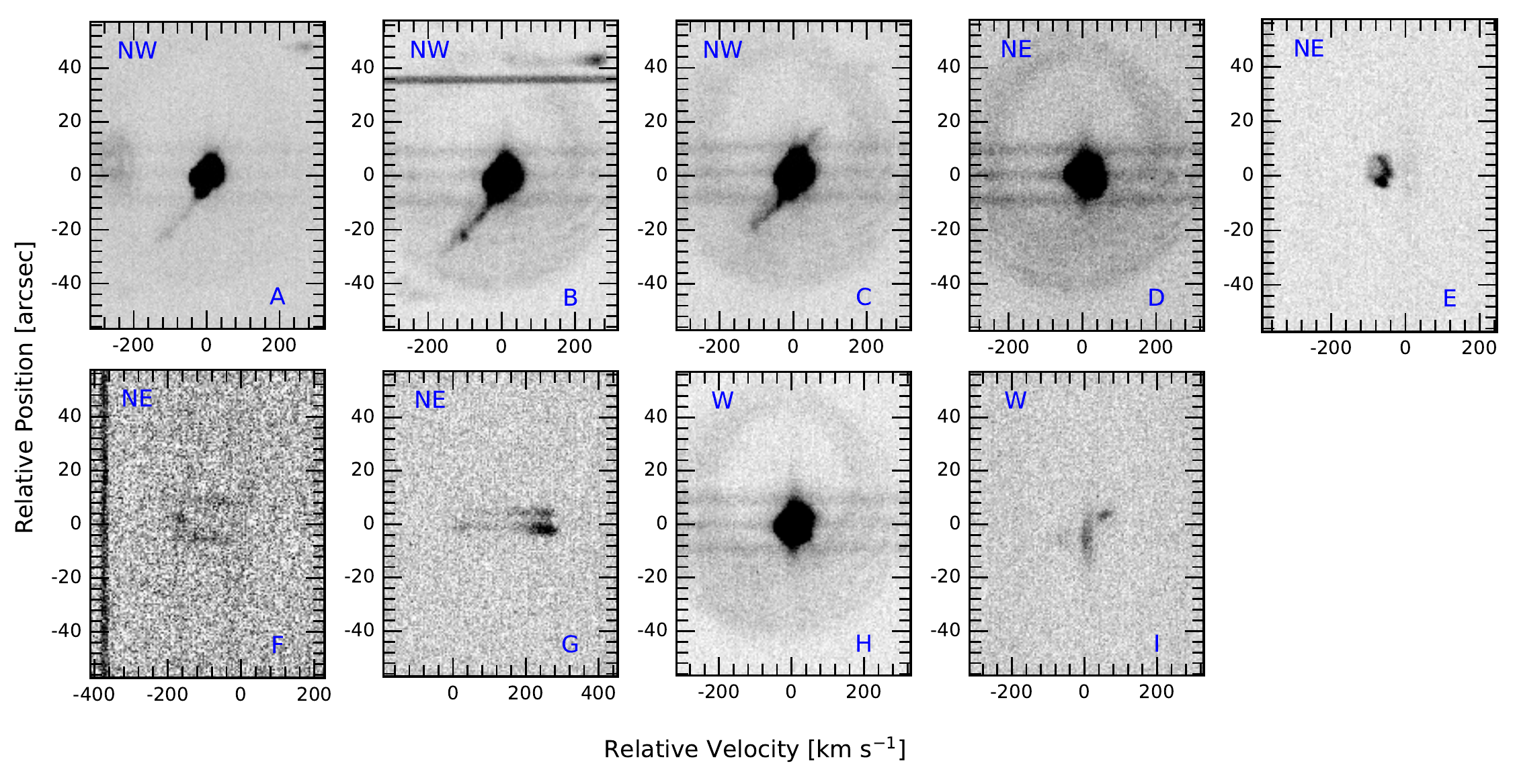}
    \caption{PV maps of the [\ion{N}{ii}] emission line for the selected slits labelled in the bottom right corner of each panel.}
    \label{fig:mez_nii_spectra}
\end{figure*}

\begin{figure*}
    \centering
    \includegraphics[width=1.0\textwidth]{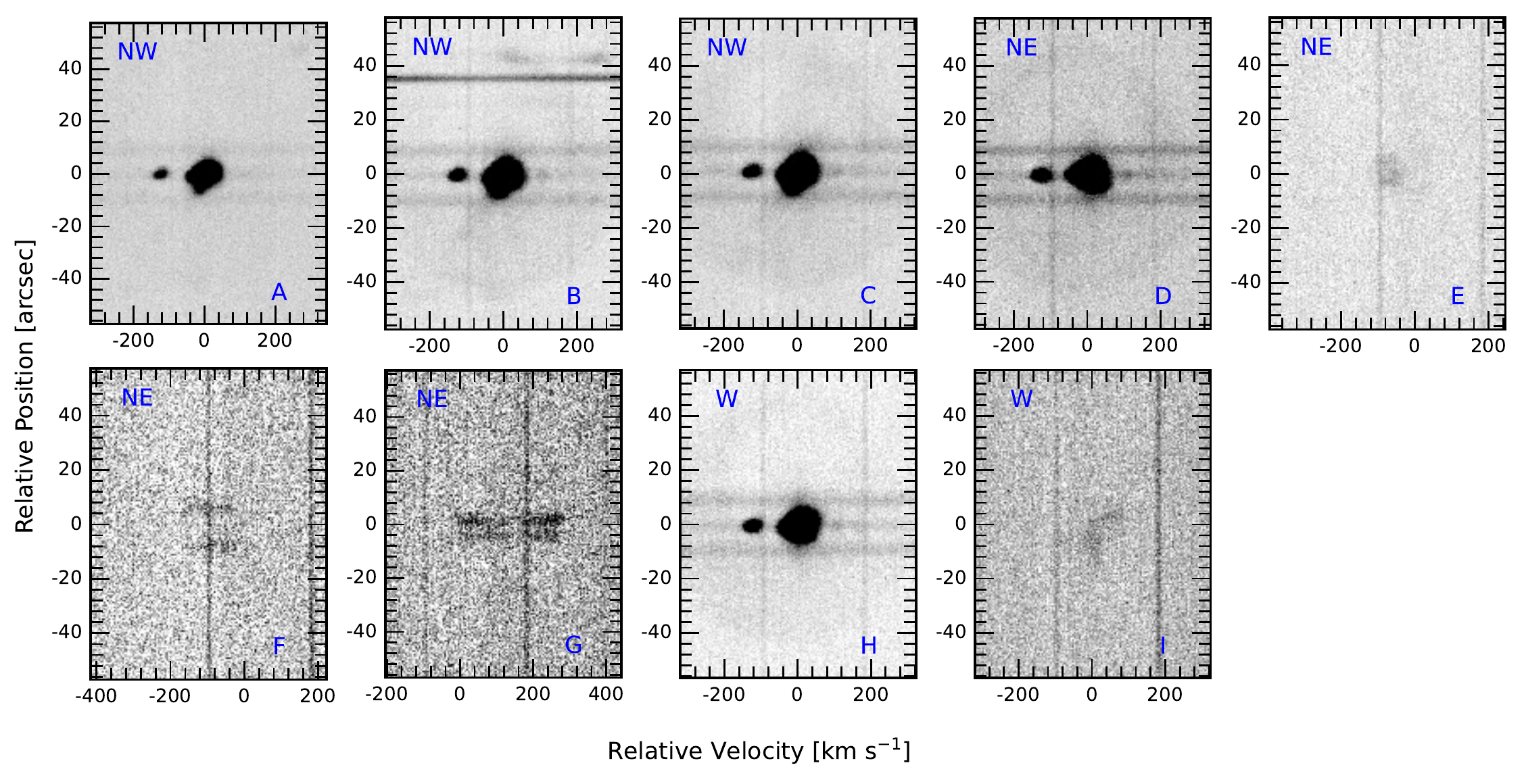}
    \caption{PV maps of the H$\alpha$ emission line for the selected slits labelled in the bottom right corner of each panel.}
    \label{fig:mez_ha_spectra}
\end{figure*}

Position-velocity maps were extracted from all the spectra. Figure~\ref{fig:mez_nii_spectra} corresponds to emission in [\ion{N}{ii}], whereas Fig.~\ref{fig:mez_ha_spectra} is for H$\alpha$. Some notable features are apparent in the panels corresponding to the slits crossing the central star (A, B, C, D, and H): (a) In Fig.~\ref{fig:mez_nii_spectra}, some artefacts can be observed, such as a `big bubble' surrounding all the emission, as well as two horizontal lines above and below the continuum of the central star. These known artefacts were briefly mentioned by \citet{meaburn2003} and examples can be found in the database of \citet{lopez2012}. (b) These artefacts are also present in the corresponding panels in Fig.~\ref{fig:mez_ha_spectra}, where the presence of the \ion{He}{ii}$\lambda$6560{\AA} emission line is also noticeable. 
Moreover, it is important to note that the emissions observed in panels F and G from both Figs.~~\ref{fig:mez_nii_spectra} and \ref{fig:mez_ha_spectra} are authentic, and they should not be confused with the artefacts previously mentioned, as they correspond to gaseous knots with low brightness.

\pagebreak

\section{The spectral unsharp masking technique}
\label{apendice_sum}

Unsharp masking is a technique used in image processing to enhance the perception of edge details in an image. It works by subtracting a blurred version of the image from the original image, accentuating the edges and fine details present in the original but not in the blurred version. The result is a sharper, more detailed image. The technique is often used in digital photography and image editing software to improve the clarity and sharpness of photographs.
In signal processing, an unsharp mask is equivalent to a filter that amplifies the high-frequency components of a signal.

The spectral unsharp masking (SUM) technique is presented here as a variant. In astronomical image processing, a blurred image is typically obtained by applying a Gaussian convolution. The main parameter of this convolution is the standard deviation or $\sigma$ value, which is related to the 'seeing' of the observations. A 1-$\sigma$ convolved image typically has no effect, as the seeing already accounts for any image blur. Commonly, a 5-$\sigma$ convolution or higher is used to obtain the blurred image. We shall call this variable $\sigma_{\rm spatial}$. 

In the case of spectra, one axis corresponds to the slit (spatial axis), and the other corresponds to the spectral axis (usually wavelength, but in our case, velocity). The spectral resolution is measured by fitting Gaussian curves to comparison arc lines. This measurement considers all types of broadenings, including those caused by local effects and instrumental errors, with natural broadening being dominant. Natural broadening also has a Gaussian profile, and we usually report the FWHM as the main parameter to describe it. \edit{FWHM is related to $\sigma$ as ${\rm FWHM}=2\sqrt{2\,\ln 2}\,\sigma$}. We shall call this variable $\sigma_{\rm spectral}$.

Since the values of $\sigma_{\rm spatial}$ and $\sigma_{\rm spectral}$ are not related to each other, and are typically different in terms of pixels, we cannot use a two-dimensional circular Gaussian, as is done in imaging. Instead, we must use a two-dimensional elliptical Gaussian. Therefore, if we want to apply a 5-$\sigma$ blur to a PV map, we must multiply the number five by the corresponding value (in pixels) to both $\sigma_{\rm spatial}$ and $\sigma_{\rm spectral}$ separately. These values are the major and minor axes of the Gaussian ellipse used to obtain the blurred image when applied to the corresponding axes. The other steps for getting the unsharp masking are the same as for imaging.

\end{appendix}

\end{document}